\documentstyle[aps,psfig,preprint,tighten]{revtex}

\begin{document}

\title{Nuclear Transparency to Intermediate-Energy Protons}
\author{James J. Kelly}
\address{ Department of Physics, University of Maryland, 
          College Park, MD 20742 }
\maketitle

\begin{abstract}
Nuclear transparency in the $(e,e^\prime p)$ reaction for 
$135 \leq T_p \leq 800$ MeV is investigated using the distorted wave
approximation.
Calculations using density-dependent effective interactions, both empirical
and theoretical, are compared with phenomenological optical potentials.
We find that nuclear transparency is well correlated with proton absorption
and neutron total cross sections and that calculations using 
density-dependent effective interactions provide
the best agreement with data.
Nuclear transparency calculations are compared with recent electron 
scattering data for $Q^2 < 2$ (GeV/c)$^2$.
For $T_p \lesssim 200$ MeV we find that there is considerable sensitivity 
to the choice of optical model and that the empirical effective interaction
provides the best agreement with the data, but remains 5--10\% low.
For $T_p \gtrsim 300$ MeV we find that there is much less difference between
these models, but that the calculations significantly underpredict
transparency data and that the discrepancy increases with $A$.
The differences between Glauber and optical model calculations are related
to their respective definitions of the semi-inclusive cross section.
By using a more inclusive summation over final states the Glauber model
emphasizes nucleon-nucleon inelasticity, whereas with a more restrictive
summation the optical model emphasizes nucleon-nucleus inelasticity;
experimental definitions of the semi-inclusive cross section lie between 
these extremes. 
\end{abstract}
\pacs{25.30.Dh,24.10.Ht,25.40-h}

\section{Introduction}
\label{sec:intro}

The transparency of nuclear matter to the propagation of an intermediate-energy
nucleon is of fundamental importance to the interpretation of many nuclear
reactions.
The elastic scattering of intermediate-energy nucleons, here defined by the
range $100 \lesssim T_p \lesssim 800$ MeV, is usually described using a 
complex optical potential whose real part produces refraction and whose 
imaginary part produces absorption of the flux contained within the elastic 
channel.
That flux reappears, of course, in other channels which include inelastic
scattering, knockout, and other reactions.
Phenomenological optical potentials are obtained by fitting the parameters
of some hypothetical function to elastic scattering data; sometimes absorption
or total cross section data are included in the analysis also.
Excellent fits to the data for a particular target and energy can usually
be achieved, but the fitted parameters often vary erratically with respect
to either mass or energy.
Global analyses of more extensive data sets which impose smooth dependencies 
on energy and/or mass, usually at the expense of local fit quality, 
are expected to produce more realistic optical potentials.
The optical potential is then presumed to represent the nucleon wave function
within the medium with sufficient accuracy for the analysis of other reactions
involving one or more nucleons of similar energy in either the initial or the
final channel, such as $(e,e^\prime N)$ or $(p,p^\prime N)$.

However, since elastic scattering is determined by asymptotic phase shifts,
any potential which produces the same set of phase shifts will predict the 
same elastic scattering even if the interior wave functions are quite 
different.  
Thus, various phase-equivalent optical potentials can produce significantly
different predictions when employed to analyze other reactions.
Alternatively, we have developed an empirical effective interaction (EEI) 
\cite{Kelly89a,Kelly89b,Kelly90a,Kelly90b,Kelly91a,Flanders91}  
based upon the density dependence predicted by nuclear matter theories
of the effective interaction, but where the parameters are fitted to
inelastic scattering data.
Transition densities fitted to electroexcitation data are used to minimize
uncertainties due to nuclear structure.
Both elastic and inelastic potentials are obtained by folding the same
density-dependent effective interaction with transition densities using the
local density approximation.
Since both the distorted waves and the transition potentials depend upon the
same interaction, the fitting procedure involves a self-consistency cycle
which converges quite quickly. 
Several inelastic transitions, possibly among several targets, are fitted 
simultaneously, producing an empirical effective interaction (EEI) that 
depends upon local density but which is independent of target.
Elastic scattering data can be analyzed also, but the interaction fitted to
inelastic scattering data usually produces good elastic predictions whether 
or not elastic data are included.

	Basing the phenomenology of the effective interaction primarily upon
inelastic scattering offers several advantages over the determination
of the optical potential via elastic scattering.  
First, the various radial shapes of transition densities provide differential 
sensitivity to the density dependence of the effective interaction, 
whereas the elastic optical potential requires only a global average over 
density.  
Second, since the distorted waves are determined self-consistently using 
optical potentials constructed from the same interaction that drives the 
inelastic transitions, the inelastic observables depend upon overlap integrals 
and are sensitive to the wave functions in the nuclear interior, 
thereby helping to determine the interior optical potential.  
Elastic scattering, on the other hand, depends on
asymptotic phase shifts and all potentials which share the same
asymptotic wave functions predict the same elastic scattering despite
differences in the interior.  
Therefore, provided that a consistent description of both elastic and 
inelastic scattering emerges from the phenomenological analysis of the 
effective interaction, the resultant optical potential should represent a 
more realistic description of the nuclear interior than models restricted 
to simple geometries, even if the latter yield  better $\chi^2$ fits to 
the elastic scattering data alone.  
Finally, the EEI model requires far fewer parameters than traditional
global optical models.

Single-nucleon knockout by electron scattering, $(e,e^\prime N)$ reactions,
provide important tests for models of nucleon propagation.
Since the nucleus is practically transparent with respect to the electron
beam, the ejectile can originate from anywhere within the nuclear volume.
Similarly, the smaller number of strongly absorbed wave functions gives 
$(e,e^\prime N)$ reactions better sensitivity to the interior than 
$(p,N)$ or $(p,p^\prime N)$ reactions.
Hence, $(e,e^\prime p)$ reactions view the optical potential quite differently
from $(p,p^\prime)$ reactions and 
there is no guarantee that the phenomenological potential that provides the
best fit to proton elastic scattering data will also provide the best 
description of $(e,e^\prime p)$ data.

Quantitative analysis of missing-momentum distributions or spectroscopic 
factors for exclusive $A(e,e^\prime p)B$ reactions to discrete states of the 
residual nucleus depends upon accurate knowledge of absorption and distortion 
by final-state interactions (FSI) \cite{Dieperink90,Boffi93,Kelly96}.
Most experiments of that type have been performed using relatively low
ejectile energies, typically $T_p \lesssim 135$ MeV, where the EEI method
is less successful.
More recent experiments at Mainz, with $T_p \approx 200$ MeV, and future
experiments at CEBAF should be less sensitive to FSI uncertainties.
Nevertheless, it is important to test those models of nuclear transparency
directly.

The first measurements of nuclear transparency using electron scattering
were made at MIT by Garino {\it et al.} \cite{Garino92,Geesaman89}
for $T_p \approx 180$ MeV by comparing inclusive cross sections with
measurements of single-nucleon knockout cross sections that sampled the 
reaction cone and which were integrated with respect to missing energy.
These semi-inclusive $(e,e^\prime p)$ data were analyzed using a correlated 
Glauber approximation by Pandharipande and Pieper \cite{Pandharipande92}, 
who found that Pauli blocking and short-range correlations are required to 
obtain transparencies large enough to reproduce the data.
However, the Glauber model \cite{Glauber59} is a high-energy approximation 
which postulates linear trajectories and hence is not expected to be 
particularly accurate for $T_p \approx 180$ MeV.
For proton elastic scattering the Glauber model is generally considered 
adequate for $T_p \gtrsim 800$ MeV \cite{Bassichis71,Frankfurt94}.
Although integrated quantities, such as nuclear transparency, are probably
less sensitive to the details of final-state interactions such that the
eikonal approximation may be sufficiently accurate at lower ejectile energies,
the lower limit of the Glauber approximation to $(e,e^\prime p)$ has not
yet been established.

At high $Q^2$, Brodsky \cite{Brodsky82} and Mueller \cite{Mueller82} have
predicted that nuclear transparency might be significantly enhanced by
the phenomenon of color transparency (CT), in which a Fock component of the 
nucleon wave function that is smaller in configuration space than the
complete nucleon is ejected as an effectively color-neutral object that 
propagates with reduced interactions and increased attenuation length.
Early calculations based upon the Glauber approximation with energy-independent
nucleon-nucleon cross sections and various models of the hadron formation 
length for point-like configurations predicted substantial enhancement of 
nuclear transparency for the $A(e,e^\prime p)$ reaction at high $Q^2$
\cite{Farrar88,Frankfurt90,Jennings92,Benhar92}.
Nuclear transparency data using the $A(e,e^\prime p)$ reaction
for $1.0 \leq Q^2 \leq 6.8$ (GeV/c)$^2$ have recently been obtained at SLAC 
by the NE18 collaboration \cite{Makins94,O'Neill95},
but no definitive signal for CT was discerned \cite{Nikolaev94b}.
On the other hand, the data were limited by statistics and new data of much
better accuracy are expected from CEBAF soon, although the maximum $Q^2$ will
be smaller.
However, more refined calculations, including the energy-dependence of the
nucleon-nucleon interaction \cite{Frankfurt94}, finite range effects
\cite{Nikolaev93a}, Fermi motion \cite{Kohama93,Nikolaev93b},
and coherency constraints \cite{Nikolaev94a},
suggest that the onset of color transparency is rather slow and cannot be
seen in these types of experiments unless $Q^2 \gtrsim 30$ (GeV/c)$^2$.
Hence, the CT signal in semi-inclusive $A(e,e^\prime p)$ cross sections is 
expected to be small at SLAC kinematics and even smaller at CEBAF kinematics. 
It has been suggested that for discrete states of the residual
nucleus the attenuation coefficient for single nucleon knockout
\cite{Frankfurt90} or the asymmetry between missing momenta parallel versus
antiparallel to the momentum transfer \cite{Jennings93} might be more
sensitive to color transparency at intermediate energies, 
but recent studies of the accuracy of the Glauber approximation suggest 
that $Q^2 \gtrsim 2$ (GeV/c)$^2$ would be required to employ that model
\cite{Nikolaev95a}.

In this paper we report calculations of nuclear transparency for
$135 \leq T_p \leq 800$ MeV using a distorted wave approximation (DWA).
We compare calculations based upon the EEI model and a relativistic
effective interaction, known as IA2 \cite{Furnstahl93,Kelly94a}, 
with a global optical potential from Dirac phenomenology \cite{Cooper93}.
For low energies we also consider several traditional nonrelativistic
optical models.
The calculations are compared with the MIT data \cite{Garino92,Geesaman89}
and with the low-energy data from SLAC experiment NE18 
\cite{Makins94,O'Neill95}.
The model is presented in Sec.\ \ref{sec:model}.
Predictions of proton absorption and neutron total cross sections using
these optical potentials are compared with the available data for these
closely related quantities in Sec.\ \ref{sec:comparison-of-potentials}.
The DWA results for nuclear transparency are presented in 
Sec.\ \ref{sec:results}.
In Sec.\ \ref{sec:discussion} we compare our results with those obtained
using the Glauber model.
We find that the difference between these approaches arises from different
definitions of the semi-inclusive cross section, with Glauber model being
more inclusive and the optical model less inclusive than the experimental
semi-inclusive cross section for intermediate-energy ejectiles.
Finally, our conclusions are summarized in Sec.\ \ref{sec:conclusions}.

\section{Model}
\label{sec:model}

\subsection{Definition of Nuclear Transparency}
\label{sec:definition}

	The distorted spectral function is obtained experimentally by
dividing the differential cross section by the off-shell electron-proton 
cross section, $\sigma_{ep}$, according to the ansatz
\begin{equation}
\label{eq:SD}
  \frac{d\sigma}{d\varepsilon_f d\Omega_e d\varepsilon_p d\Omega_p} 
              = K \sigma_{ep} S^D(E_m,{\bf p}_m, {\bf p}^\prime) \; ,
\end{equation}
where $K$ is a kinematical factor.
Final-state interactions between the ejectile and the residual nucleus
make the distorted spectral function $S^D(E_m,{\bf p}_m, {\bf p}^\prime)$
depend upon the ejectile momentum ${\bf p}^\prime$, and on the angle between
the initial and final nucleon momenta, whereas the (undistorted) spectral 
function would depend only on $E_m$ and $p_m$.
Thus, the distorted spectral function depends upon the kinematical 
conditions and is different for parallel and perpendicular kinematics, 
for example. 
Furthermore, the dependence of $S^D$ upon the electron energy that arises
from Coulomb distortion and from the properties of the electron current
has been left implicit.

Due to the distribution of initial momenta, the quasifree single-nucleon
knockout strength is spread over a Fermi cone whose opening angle is
approximately $\theta_F = \tan^{-1} k_F/q$ and over a range of missing
energies that includes the binding energy for the deepest orbital.
However, final-state interactions broaden these distributions and transfer
some of the flux into more complicated final states.
Hence, nuclear transparency can be loosely defined as the ratio between 
the coincident $(e,e^\prime p)$ cross section and the inclusive quasifree
electron scattering cross section, where the concidence cross section is
integrated over the Fermi cone and over the range of missing energy 
populated by direct knockout.

Optical models describe the loss of flux from the initial single-nucleon 
channels as absorption.
Hence nuclear transparency is calculated in the optical model as
the ratio
\begin{equation}
\label{eq:transparency}
   {\cal T}_w = 
  \frac{ \int dE_m \int d^3 p_m \; w(E_m,{\bf p}_m, {\bf p}^\prime)
				   S^D(E_m,{\bf p}_m, {\bf p}^\prime) }
       { \int dE_m \int d^3 p_m \; w(E_m,{\bf p}_m, {\bf p}^\prime)
				   S^{PW}(E_m, p_m) }
\end{equation}
where in the numerator the distorted spectral distribution includes 
final-state interactions, and depends upon the ejectile momentum, 
whereas in the denominator the undistorted spectral distribution 
depends only upon the missing energy and momentum.
The weight factor, $w(E_m,{\bf p}_m, {\bf p}^\prime)$, represents the
experimental acceptance and distinguishes, for example, between parallel
or quasiperpendicular kinematics; hence we distinguish various types of
transparency functions using a subscript $w$ signifying the appropriate
acceptance function.
Clearly distortion and transparency depend upon the ejectile energy.
However, since the distorted spectral function must be evaluated for the 
appropriate kinematical conditions, its dependence upon electron-scattering 
kinematics remains implicit in this definition.
Therefore, to complete the definition of transparency we would need to
specify the kinematical conditions of interest more completely, using the
same integration regions and weighting factors for both numerator and 
denominator, 
where the appropriate weighting factors depend upon the kinematics of the
experiment.

The simplest situation arises when $(\omega,q)$ are held constant, for which
the nuclear transparency ${\cal T}_\bot$ may be defined as
\begin{equation}
\label{eq:transparency-perp}
   {\cal T}_\bot = 
  \frac{ \int dE_m \int d\theta \sin\theta \; 
         S^D(E_m, p_m, p^\prime, \theta) }
       { \int dE_m \int d\theta \sin\theta \; S^{PW}(E_m,p_m) }
\end{equation}
where $\theta$ is the angle between the ejectile momentum ${\bf p}^\prime$
and the momentum transfer ${\bf q}$.
Furthermore, it is simplest to require that the ejectile kinetic energy
be constant in the barycentric frame so that final--state interactions can
be evaluated for a unique ejectile energy.
Also note that the weight factor should be expressed in terms of the
center-of-mass system, where the ejectile momentum is constant, but
for large $q$ the laboratory momentum changes little over the Fermi
cone so that the distinction between the lab and cm angles matters
little.
However, for experiments using a narrow acceptance in $\omega$ with
a large acceptance in ejectile kinetic energy $T_p$, 
it is necessary to employ energy--dependent optical potentials.
Alternatively, for parallel kinematics we define
\begin{equation}
\label{eq:transparency-para}
   {\cal T}_\| = 
  \frac{ \int dE_m \int dp_m \, p_m^2 \; S^D(E_m,p_m,p^\prime) }
       { \int dE_m \int dp_m \, p_m^2 \; S^{PW}(E_m,p_m) }
\end{equation}
but must recognize that the ejectile momentum $p^\prime = p_m + q$
is correlated with $p_m$ and that the electron scattering kinematics 
again remain implicit.
In this case it is also clearest to require the invariant mass for each
$E_m$ to be constant despite the concomitant variation of electron
kinematics.

We have verified that when the optical potentials are nullified, the
distorted wave calculations result in unit transparency.
However, it is important to recognize that these definitions of nuclear 
transparency can produce deviations from unity, with either sign, 
even when the optical potential is purely real because weighting functions 
based upon the plane-wave impulse approximation do not account for refractive 
effects.
Although one might be tempted to divide out the refractive effect of the 
real potential by using a modified definition
\begin{equation}
   {\tilde {\cal T}}_w = 
  \frac{ \int dE_m \int d^3 p_m \; w(E_m,{\bf p}_m, {\bf p}^\prime)
				   S^D(E_m, {\bf p}_m, {\bf p}^\prime) }
       { \int dE_m \int d^3 p_m \; w(E_m,{\bf p}_m, {\bf p}^\prime)
				   S^R(E_m, {\bf p}_m, {\bf p}^\prime) }
\end{equation}
where $S^R(E_m, {\bf p}_m, {\bf p}^\prime)$ is the distorted spectral
function for the real part of optical potential,
our original definition, Eq. (\ref{eq:transparency}), conforms more closely
to the customary experimental definition.
The difference between these approaches is generally greatest when the 
summation over missing energy is restricted to a single state, or subshell.
For quasiperpendicular kinematics, the refractive effects are minimized
when measurements are made on both sides of ${\bf q}$ for closed shell
targets so that the opposing effects of spin-orbit distortion are 
approximately balanced for both sides of ${\bf q}$ and for both 
spin-orbit partners.
For parallel kinematics a net attractive (repulsive) real central potential
shifts the missing momentum distributions to smaller (larger) $p_m$ and 
enhances (reduces) the peak values of $S^D(E_m,p_m)$ for positive (negative) 
missing momenta,
but the net effect on transparency tends to balance when the integration 
over $p_m$ is symmetric.
Since the shifts in the peak positions are also sensitive to spin-orbit 
distortion, the attenuation factors for spin-orbit partners can be different.
Nevertheless, numerical studies show that for calculations using symmetric
ranges of missing momenta the ratio ${\cal T}_w / {\tilde {\cal T}}_w$ 
remains within a few percent of unity for both parallel and 
quasiperpendicular kinematics
for $A \geq 12$ and $135 \leq T_p \leq 800$ MeV.
Alternatively, one can assess the effects of refraction by comparing
transparency calculations with and without the real parts of the optical
potential.
Thus, over the ranges of $A$ and $T_p$ considered, 
we find that $|\delta {\cal T}| / {\cal T} \lesssim 7\%$ when
the real parts of the optical potential are eliminated and conclude that
the present definition does in fact provide an unambiguous measure of
transparency that is rather insensitive to refraction.
Therefore, we prefer to employ Eq. (\ref{eq:transparency}), despite its
slight ambiguity between refraction and absorption, because it is closer to
the experimental quantity of interest and the refractive effects do not
significantly affect its interpretation in terms of absorption.
Furthermore, we also find that
\begin{equation}
 \left| \frac{ {\cal T}_\| - {\cal T}_\bot }{ {\cal T}_\bot } \right|
 \lesssim 6\%
\end{equation}
for $A \geq 12$ and $135 \leq T_p \leq 800$ MeV; 
hence, we expect our calculations to be quite insensitive to the small
range of missing momentum parallel to ${\bf q}$ that is inevitably accepted 
when performing measurements of ${\cal T}_\bot$.
Therefore, we compare calculations of ${\cal T}_\bot$ directly to the
experimental data without attempting to simulate the complete experimental
acceptance functions.

\subsection{Spectral Function}
\label{sec:spectral}

In the independent--particle model (IPM) the spectral functions take the
forms
\begin{eqnarray}
\label{eq:IPM}
   S_{IPM}(E_m,p_m) &=& 
    \sum_\alpha S_\alpha 
     \varrho_\alpha(p_m) \delta(E_m-E_\alpha) \\
   S^D_{IPM}(E_m,{\bf p}_m,{\bf p}^\prime) &=& 
    \sum_\alpha S_\alpha 
     \varrho^D_\alpha({\bf p}_m,{\bf p}^\prime) \delta(E_m-E_\alpha)
\end{eqnarray}
where $E_\alpha$ and $S_\alpha$ are the missing energy and spectroscopic 
factor for orbital $\alpha$,
\begin{equation}
 \varrho_\alpha(p_m) =
  \left| \int d^3r \: e^{-i{\bf p_m \cdot r}} \phi_\alpha({\bf r}) \right|^2
\end{equation}
is the momentum distribution arising from a  single-particle
wave function $\phi_\alpha({\bf r})$ normalized such that 
\begin{equation}
 4\pi \int dp_m \; p_m^2 \; \varrho_\alpha(p_m) = 1 \; ,
\end{equation}
and $\varrho^D_\alpha({\bf p}_m,{\bf p}^\prime)$ is the corresponding
distorted momentum distribution. 
Hence the nuclear transparency for quasiperpendicular kinematics reduces to
\begin{equation}
\label{eq:perp}
   {\cal T}_\bot = 
  \frac{ \sum_\alpha S_\alpha \int d\theta \sin\theta \; 
      \varrho^D_\alpha(p_m,p^\prime,\theta) }
       { \sum_\alpha S_\alpha \int d\theta \sin\theta \; 
      \varrho_\alpha(p_m) } \; .
\end{equation}

Although interactions fragment the low-lying valence hole strength and
populate the continuum for large missing energy and momentum, 
nuclear transparency is much more sensitive to final state interactions 
than to details of the spectral distribution.
For a given orbital, attenuation depends more strongly on ejectile energy
than on missing energy, so that the spreading of the hole strength is not
expected to appreciably affect its contribution to the integrated yield,
especially for experiments which select a relatively narrow range of
ejectile energy.
Similarly, although variations of the bound--state wave function which
change the rms radius affect the missing momentum distribution, such
variations have little effect on the integrated yield and tend to cancel
in the ratio used for transparency.
Therefore, for transparency calculations it is sufficient to employ IPM
spectral functions, although correlations must be included to describe the
distorted spectral distribution in detail.

For $A \leq 16$ we used Woods-Saxon single-particle wave functions, whereas
for heavier nuclei bound-state wave functions were obtained using the 
Skyrme-Hartree-Fock model based upon the interaction designated $Z_\sigma$ 
\cite{Friedrich86,Reinhard91}.
The single-particle energy spectra for each target were shifted to
obtain the correct separation energies.
For each orbital the final--state interactions were evaluated using 
ejectile energies based upon $\omega$ and the shifted Hartree-Fock
separation energies.
Note that the results are rather insensitive to the details of the
bound-state wave functions, but are quite sensitive to the choice of
optical model.

\subsection{Distorted Wave Approximation}
\label{sec:DWA}

The distorted wave approximation (DWA) for the electromagnetic transition 
amplitude that governs the single-nucleon knockout reaction
$A(e,e^\prime N)B$ can be expressed in the form \cite{deForest67}
\begin{equation}
  {\cal M} = \int \frac{d^3q^\prime}{(2\pi)^3} \;
  {\cal J}^e_\mu({\bf q}^\prime) \frac{1}{Q^{\prime 2}} 
  {\cal J}_N^\mu({\bf q}^\prime)
\end{equation}
where the electron and nuclear currents are
\begin{mathletters}
\begin{eqnarray}
 {\cal J}^e_\mu({\bf q}^\prime) &=& \int d^3r \;
        e^{-i {\bf q}^\prime \cdot {\bf r}}
   \bar{\psi}^e_f({\bf r}) \gamma_\mu \psi^e_i({\bf r}) \\
 {\cal J}^N_\mu({\bf q}^\prime) &=& \int d^3r \;
     e^{i \tilde{{\bf q}}^\prime \cdot {\bf r}}
   \bar{\psi}^N_f({\bf r}) \Gamma_\mu \psi^N_i({\bf r}) 
\end{eqnarray}
\end{mathletters}
and where $\Gamma_\mu$ is the vertex operator for the nucleon current.
In these expressions the electron wave functions relative to the
target of mass $m_A$ are denoted by the spinors
$\psi^e_i$ and $\psi^e_f$ for the initial and final states, respectively.
At this stage we leave implicit the dependence of the nuclear current upon
the ejectile kinematics and the state of the residual nucleus.
Since it is more convenient to express the ejectile wave functions $\psi^N$ 
relative to the residual nucleus of mass $m_B$, 
the radial scale is adjusted by means of the reduced momentum transfer 
\cite{Boffi79}
$\tilde{{\bf q}}^\prime = {\bf q}^\prime m_B/m_A$. 

If we assume that a virtual photon with momentum ${\bf q}^\prime$ is absorbed 
by a single nucleon with initial momentum ${\bf p}$, the nuclear current at 
position ${\bf r}$ becomes
\begin{equation}
 {\cal J}^N_\mu({\bf r}) = \int \frac{d^3p}{(2\pi)^3} 
                           \frac{d^3p^{\prime \prime}}{(2\pi)^3} \;
     e^{-i \tilde{{\bf q}}^\prime \cdot {\bf r}} 
     \tilde{\chi}^{(-)*}({\bf p}^\prime, {\bf p}^{\prime \prime}) 
     \Gamma_\mu({\bf p}^{\prime \prime}, {\bf p}) \tilde{\phi}({\bf p})
\end{equation}
where the single-nucleon wave function is the amplitude for removing a
nucleon from the initial state of target A and reaching the final state
of residual nucleus B, such that
\begin{equation}
  \tilde{\phi}({\bf p}) = \langle B | a({\bf p}) | A \rangle \; .
\end{equation}
The distorted wave 
$\tilde{\chi}^{(-)*}({\bf p}^\prime, {\bf p}^{\prime \prime})$
is the amplitude that the ejectile with initial momentum 
${\bf p}^{\prime \prime} = {\bf p} + {\bf q}^\prime$  emerges from the nuclear 
field with final momentum ${\bf p}^\prime$.
In coordinate space these wave functions are expressed as
\begin{mathletters}
\begin{eqnarray}
 \phi({\bf r}) &=& \int \frac{d^3p}{(2\pi)^3} \; 
                 e^{i {\bf p} \cdot {\bf r}} \tilde{\phi}({\bf p}) \\
 \chi({\bf p}^\prime,{\bf r}) &=& \int \frac{d^3p^{\prime \prime}}{(2\pi)^3} \; 
                 e^{i {\bf p}^{\prime \prime} \cdot {\bf r}} 
 \tilde{\chi}({\bf p}^\prime, {\bf p}^{\prime \prime}) \; .
\end{eqnarray}
\end{mathletters}
Thus, the nuclear current becomes
\begin{equation}
 {\cal J}^N_\mu({\bf p}^\prime,{\bf q}^\prime) = \int \frac{d^3p}{(2\pi)^3} \;
 \tilde{\chi}^{(-)*}({\bf p}^\prime, {\bf p}+{\bf q}^\prime)
 \Gamma_\mu({\bf p}+{\bf q}^\prime, {\bf p}) \tilde{\phi}({\bf p})
\end{equation}
where ${\bf q}^\prime$ is the local momentum transfer supplied by the electron.

Since nuclear transparency depends upon the nuclear final--state 
interactions and not the electronuclear initial--state interactions, 
distortion of the electron wave function should either be included in
both numerator and denominator of Eq.\ (\ref{eq:transparency})
or excluded from both.
To a good approximation, Coulomb distortion can be described as a shift of 
the effective momentum transfer and a focussing factor which increases the
virtual--photon flux \cite{Schiff56,Jin93b}.
Since these aspects of Coulomb distortion have similar effects upon both 
exclusive and inclusive electron scattering, it is reasonable to omit Coulomb 
distortion for both.
In the absence of Coulomb distortion, the electron current is proportional to 
$\delta^3({\bf q}^\prime - {\bf q})$, so that the nuclear current can be 
evaluated for a unique value of the momentum transfer obtained from asymptotic 
kinematics.
Therefore, we obtain
\begin{equation}
\label{eq:JN-EMA}
 {\cal J}^N_\mu({\bf p}^\prime, {\bf q}) \approx \int d^3r \; 
             e^{i \tilde{{\bf q}} \cdot {\bf r}}
             \chi^{(-)*}({\bf p}^\prime, {\bf r})
             \Gamma_\mu({\bf p}^\prime, {\bf p}^\prime - {\bf q})
             \phi({\bf r}) 
\end{equation}
where the vertex function has now been reduced to a matrix, acting on nucleon
spins, whose elements are evaluated using effective kinematics.

We used the {\it cc1} vertex function of de Forest \cite{deForest83} with
nucleon form factors from model 3 of Gari and Kr\"umpelmann
\cite{Gari92a,Gari92b}.
Current conservation was enforced at the one-body level by eliminating the
longitudinal in favor of the charge component.
However, it is important to note that the transparency calculations are
quite insensitive to these choices for the vertex function and, in fact,
are quite insensitive to the electron--scattering kinematics also.
The most important variables are the ejectile energy and the choice of optical
model.

\subsection{Optical Models}
\label{sec:potentials}

Distorted waves were obtained from solutions to a Schr\"odinger equation of 
the form
\begin{equation}
  (\nabla^2 + k_\alpha^2 - 2\mu_\alpha U_\alpha) 
  \xi_{\alpha}(\bbox{k}_\alpha,\bbox{r}) = 0
\end{equation}
where relativistic kinematics are incorporated by interpreting $k_\alpha$ as
the exact relativistic wave number and $\mu_\alpha$ as the reduced energy
for channel $\alpha$ \cite{Ray92}.
We assume that the optical potential can be reduced to local form and that
nonspherical components may be neglected.
Thus, the optical potential takes the form
\begin{equation}
   U_\alpha(r) = U_\alpha^Z(r) + U_\alpha^C(r) + 
   U_\alpha^{LS}(r) {\bf L} \cdot \bbox{\sigma} 
\end{equation}
where $U_\alpha^Z$ is the Coulomb potential, $U_\alpha^C$ is the central 
potential, and 
$U_\alpha^{LS} = {\displaystyle \frac{1}{r} } 
                 {\displaystyle \frac{\partial F^{LS}}{\partial r} }$ 
is the spin-orbit potential for exit channel $\alpha$.  
Although in principle the optical potential for each exit channel depends upon
the structure of the residual nucleus, we employ a mean field approximation 
for single-nucleon knockout in which there is only a small kinematic
dependence upon the ejectile energy.
To accomodate models which include nonlocality corrections, we identify
the distorted wave function as
\begin{equation}
 \chi_{\alpha}(\bbox{k}_\alpha,\bbox{r}) = P(r) 
  \xi_{\alpha}(\bbox{k}_\alpha,\bbox{r})
\end{equation}
where $P(r)$ is a Perey factor which is unity for local models, approaches
unity at large distances for nonlocal models, and which may be complex.

\subsubsection{Nonrelativistic Woods-Saxon Potentials}

The potential fitted by Schwandt {\it et al}.\ \cite{Schwandt} to cross
section and analyzing power for proton elastic scattering for $A \geq 40$
and $80 \leq T_p \leq 180$ MeV is commonly employed for knockout analyses.
It may be used for either proton or neutron scattering since it contains
a parametrization of the symmetry potential.
Although the Schwandt potential does not bear extrapolation in either mass 
or energy well, it is nevertheless often used for lighter nuclei, 
sometimes even for mass-12. 
A potential developed by Abdul-Jalil and Jackson \cite{Abdul-Jalil79,Jackson80}
for $A \approx 12$ and $50 \leq T_p \leq 160$ MeV has sometimes been used
for knockout studies, but in our opinion its description of proton scattering 
data is unsatisfactory.
Alternatively, the $p+^{12}$C potential of Comfort and Karp \cite{Comfort80}
for $T_p \leq 185$ MeV is preferred.
Unfortunately, a global nonrelativistic optical potential for $A \geq 12$
and a broad range of energy does not appear to exist.
For that we must appeal to Dirac phenomenology (DP).

	Many analyses of single-nucleon knockout also include a Perey 
nonlocality correction of the form \cite{Perey62,Perey63}
\begin{equation}
\label{eq:PB}
  P(r) = \left[ 1 - \frac{m_p}{2\hbar^2} \beta^2 V^C(r) \right]^{-1/2}  
\end{equation}
where the central potential is separated into real and imaginary parts denoted 
$U^C(r) = V^C(r) + i W^C(r)$.
The nonlocality parameter is typically chosen as $\beta = 0.85$ fm, based
upon the original analysis of neutron scattering for $T_n \leq 20$ MeV.
Although the applicability of this simple prescription has not been
established for $T_p > 100$ MeV, 
when using nonrelativistic Woods-Saxon potentials, such as the Schwandt or
the Comfort and Karp models,  
we conform to standard practice by including the Perey factor.
Since the Perey factor is less than unity in the interior and equal to
unity outside the potential, its effect is to reduce the transparency,
especially for interior orbitals.

\subsubsection{Dirac Phenomenology}
\label{sec:DP}

Suppose that a four-component Dirac spinor,
\begin{displaymath}
  \Psi( {\bf r} ) = \left( 
\begin{array}{c} \psi_+({\bf r}) \\ \psi_-({\bf r}) \end{array}
                    \right)
\end{displaymath}
where $\psi_+$ and $\psi_-$ are two-component Pauli spinors for positive
and negative energy components,
satisfies a Dirac equation of the form
\begin{equation}
\label{eq:Dirac}
 \left[ \bbox{\alpha} \cdot \bbox{p} + \beta(m+S) \right] \Psi
 = (E-V-V^Z)\Psi
\end{equation}
with scalar and vector potentials $S$ and $V$.
Upon elimination of the lower components, an equivalent Schr\"{o}dinger
equation of the form 
\begin{equation}
 \left[ \nabla^2 + k^2 - 2\mu \left( U^Z + U^C + 
   U^{LS} \bbox{L} \cdot \bbox{\sigma} \right) \right] \phi = 0 
\end{equation}
can be obtained, where $\psi_+$ is related to $\phi$ by the
Darwin transformation 
\begin{mathletters}
\label{eq:Darwin}
\begin{eqnarray}
 \psi_+ &=& B^{1/2} \phi \\
 B &=& 1 + \frac{S-V-V^Z}{E+m}  \; .
\end{eqnarray}
\end{mathletters}
The Schr\"{o}dinger solutions are phase-equivalent to the Dirac solutions
in the sense that the asymptotic phase shifts, and hence observables for
elastic scattering, are the same.
However, the Dirac wave function is modified in the interior by a
nonlocality factor similar in form to the Perey-Buck nonlocality factor,  
except that it depends upon $S-V$, which is closely related to the spin-orbit
potential, rather than upon the central potential.
Thus, $B$ can be deduced directly from the spin-orbit potential  
\cite{Jin94a}.
When used in nonrelativistic calculations, the positive energy spinor $\psi_+$,
including the Darwin factor, is identified with the distorted wave $\chi$.

Hama {\it et al}.\ \cite{Hama90} produced global Dirac optical potentials
for $A \geq 40$ and $65 \leq T_p \leq 1040$ MeV.
The global Dirac optical potential was then extended by 
Cooper {\it et al}.\ \cite{Cooper93}
to the ranges $A \geq 12$ and $20 \leq T_p \leq 1040$ MeV.
Of the several essentially equivalent variations of the global potential that
were provided, we have chosen the version labelled EDAD1.
Although the scalar+vector (SV) model of Dirac phenomenolgy is not
unique, and relatively simple but arbitrary shapes are employed for the
potentials, this work represents the most extensive and systematic analyses 
of proton optical potentials available.
The available proton-nucleus elastic scattering data are described very 
well by potentials whose properties vary smoothly with both mass and energy.

\subsection{Local Density Approximation}
\label{sec:LDA}

Optical potentials can also be obtained by folding the nucleon-nucleon
effective interaction with the nuclear density distribution.
In recent years it has become clear that the intermediate-energy 
nucleon-nucleon effective interaction depends strongly upon the density in 
the interaction region.  
Several calculations of the effective interaction in nuclear matter have been 
made following the seminal work of H\"{u}fner and Mahaux  \cite{Hufner}.
Jeukenne, Lejeune, and Mahaux \cite{JLM74,JLM76,JLM77a,JLM77b}
computed the self-energy and the optical potential for $0 \leq T_p \leq
160$ MeV using the Reid soft-core potential \cite{Reid}.  
Brieva, Rook, and von Geramb 
\cite{Brieva77a,Brieva77b,Brieva78a,Brieva78b,Geramb79}
developed a Brueckner--Hartree--Fock (BHF) approach and
used the Hamada-Johnston potential \cite{HJ} to calculate the pair wave 
function in nuclear matter, from which a local pseudopotential was constructed
for $T_p \leq 180$ MeV using a generalization of the Siemens averaging 
procedure \cite{Siemens}.  
Similar calculations using the Hamada-Johnston potential have also been
performed by Yamaguchi, Nagata, and Michiyama (YNM) \cite{YNM83,YNM86},
who parametrized their results for $T_p \leq 200$ MeV in Gaussian rather 
than Yukawa form.
The BHF approach was refined by von Geramb and collaborators 
\cite{Rikus84,Geramb83},
who constructed an effective interaction based upon the 
Paris potential \cite{Lacombe80}, designated Paris-Hamburg (PH), that
is applicable for $100 \leq T_p \leq 400$ MeV.
Nakayama and Love \cite{Nakayama88}
used the Bonn potential \cite{Machleidt87} to calculate a local pseudopotential 
that reproduces on-shell matrix elements of the G-matrix.  
These theories are all based upon the Bethe-Goldstone equation and 
include Pauli blocking and self-energy corrections self-consistently.  
A closely related calculation by Ray \cite{Ray90}
used a coupled channels nucleon-isobar model and Watson multiple 
scattering theory to calculate a density-dependent t-matrix 
applicable to nucleon energies above 200 MeV that includes some of the
effects of pion production.  
Finally, Furnstahl, Wallace, and Kelly \cite{Furnstahl93,Kelly94a}
have developed an effective interaction in a form similar to the EEI model
based upon the relativistic IA2 model.
Density dependence arises from the distortion of Dirac spinors in the
nuclear medium, which primarily affects the real central interaction, and 
Pauli blocking, which damps the absorptive potential.
The model naturally provides stronger density dependence for inelastic
scattering than for elastic scattering, which is needed to describe the
data.

All of these calculations predict strong density dependence of the 
nucleon-nucleon effective interaction.  
The dominant effect for the real central interaction is equivalent to a
short--ranged repulsive interaction that is proportional to density and 
nearly independent of energy.
In the BHF approaches this short--range repulsive interaction arises both 
from the anticorrelation between identical nucleons in nuclear matter and
from dispersive effects in the self-consistent mean field, 
whereas in the IA2 approach it arises from spinor distortion in the strong 
scalar and vector mean fields.
However, both approaches predict that the density dependence of the
real central interaction depends slowly upon energy and remains quite
strong even at 800 MeV \cite{Ray92,Kelly94a}.
The dominant effect for the imaginary central interaction arises from
Pauli blocking and for both the BHF and the IA2 models gives results similar 
to the familiar Clementel--Villi \cite{Clementel}
damping of the absorptive potential, 
in which the damping factor is inversely proportional to the proton energy.
Hence this effect is most important for low energies.

	Although the qualitative features of the medium modifications
are essentially the same, the quantitative differences among the
various theories are surprisingly large, much larger than would be
expected from the variations among the underlying nucleon-nucleon 
potentials that are employed, suggesting that the approximations 
required to evaluate the effective interaction are not yet under
good control.  The effect of these differences upon elastic and inelastic
scattering calculations has been surveyed in a series of papers
by Kelly and collaborators in which transition densities measured
by electroexcitation are used to minimize uncertainties due to 
nuclear structure and to isolate the effective interaction for detailed
examination
\cite{Kelly89a,Kelly89b,Kelly90a,Kelly90b,Kelly91a,Flanders91}.  
Transition densities which are strong in the interior
provide information about the high-density properties of the effective 
interaction, whereas surface-peaked transition densities reveal the
low-density properties.  The systematic comparison of such cases
demonstrates quite clearly that the effective interaction depends upon
local density and that estimates based upon nuclear matter theory have
qualitatively correct characteristics, but that none of the theories
presently available is sufficiently accurate for quantitative 
applications to nuclear structure.  
Therefore, an empirical model of the effective interaction was developed in 
which medium modifications similar to those predicted by nuclear matter theory
are parametrized in a form suitable to phenomenological analysis of data.
The parameters are adjusted to reproduce inelastic scattering data for
several states in one or more targets simultaneously.  Empirical effective
interactions have been extracted from data for several energies in the
range $100 \leq T_p \leq 650$ MeV.  For each energy we find that a
unique effective interaction describes data for several inelastic
transitions in a single nucleus and that the fitted interaction is
essentially independent of target.  These findings tend to confirm the basic
hypothesis of the local density approximation (LDA), namely that the
interaction depends primarily upon local density and is independent
of the detailed structure of any particular target or transition.  
The fitted parameters also exhibit a relatively smooth energy dependence.

Although slightly better fits to some of the data sets may be found in the 
original analyses, for the present purposes we choose to employ the empirical
effective interactions (EEI) tabulated by Kelly and Wallace \cite{Kelly94a}.
For this set of interactions, the medium modifications are applied to
the Franey-Love (FL) \cite{FL85} parametrization of the free $t$-matrix, which
is available for all relevant energies, and common fitting strategies and 
constraints were used to help smooth the energy dependence of the fitted
parameters.
These choices are made primarily for aesthetic reasons and have very little
effect upon knockout calculations.

In the local density approximation (LDA), the central and spin-orbit 
potentials become
\begin{mathletters}
\begin{eqnarray}
   U^C(r) & = & \frac{2}{\pi} \int dq \ q^2 j_0(qr) 
                  t^C(q,\rho) \tilde{\rho}(q) \\
   F^{LS}(r) & = & \frac{2}{\pi} \int dq \ q^2 j_0(qr) 
                  t^{\prime LS}(q,\rho) \tilde{\rho}(q) 
\end{eqnarray}
\end{mathletters}
where 
\begin{equation}
   \tilde{\rho}(q) = \int dr \ r^2 j_0(qr) \rho(r) 
\end{equation}
is the Fourier transform of the ground-state density $\rho$.
Note that a sum over nucleon or isospin indices has been left
implicit.
To minimize uncertainties due to the nuclear density, the
proton density was obtained by unfolding the proton charge form
factor from the charge densities measured by electron scattering and
tabulated in Refs.\ \cite{deVries87}.  
For relatively small momentum transfers, 
charge symmetry ensures that the neutron and proton densities for 
mirror nuclei are very nearly proportional to each other and since the
high $q$ properties of distorting potentials have very little effect upon
knockout calculations, especially for integrated strengths, we use
$\rho_n(r) = N \rho_p(r) / Z$.
Furthermore, we evaluate the local density at the site of the projectile.

Since the empirical effective interaction was fitted to data for elastic
and inelastic scattering self-consistently without explicit use of a 
nonlocality correction, no Perey factor is used with the EEI model for
knockout.
Similarly, the IA2 interaction is local by construction and does not require
a Perey factor either.
Some other versions of the LDA do involve nonlocality corrections based
upon the exchange contribution or upon the momentum dependence of the 
effective mass, but are not employed here because none of those models
provide adequate descriptions of the proton scattering data.

	The potentials which emerge from all microscopic models exhibit
much more complicated radial shapes than posited by the Woods-Saxon
model of the optical potential.  The detailed shapes depend upon the
density dependence and range of the effective interaction and
upon the nuclear density, which especially for light targets is not 
well approximated by the Fermi shape.  For energies between about 100
and 300 MeV, for example, the real central potential exhibits a
characteristic "wine-bottle" shape.  Similar shapes also arise from
the nonrelativistic reduction of either Dirac phenomenolgy or the 
relativistic impulse approximation.  Although good fits to elastic 
scattering data may be achieved with simplistic models of the potential, 
artificially simple geometries cannot be justified on more fundamental 
grounds. 
Furthermore, the missing momentum distributions for discrete states
do show some sensitivity to the shape of the real central potential.

\section{Comparison of Optical Models}
\label{sec:comparison-of-potentials}

Integrated quantities, such as nuclear transparency, are much more
sensitive to attenuation of the flux than to distortion of the angular 
distribution by final state interactions.
Therefore, in this section we compare data for proton absorption 
and neutron total cross sections with calculations based upon a variety
of optical models for $100 < T_p < 800$ MeV.
More detailed analyses of proton elastic and inelastic scattering data
can be found in Refs. \cite{Kelly89a,Kelly89b,Kelly90a,Kelly90b,Kelly91a},
for example.  

\begin{figure}[htbp]
\centerline{
\strut\psfig{file=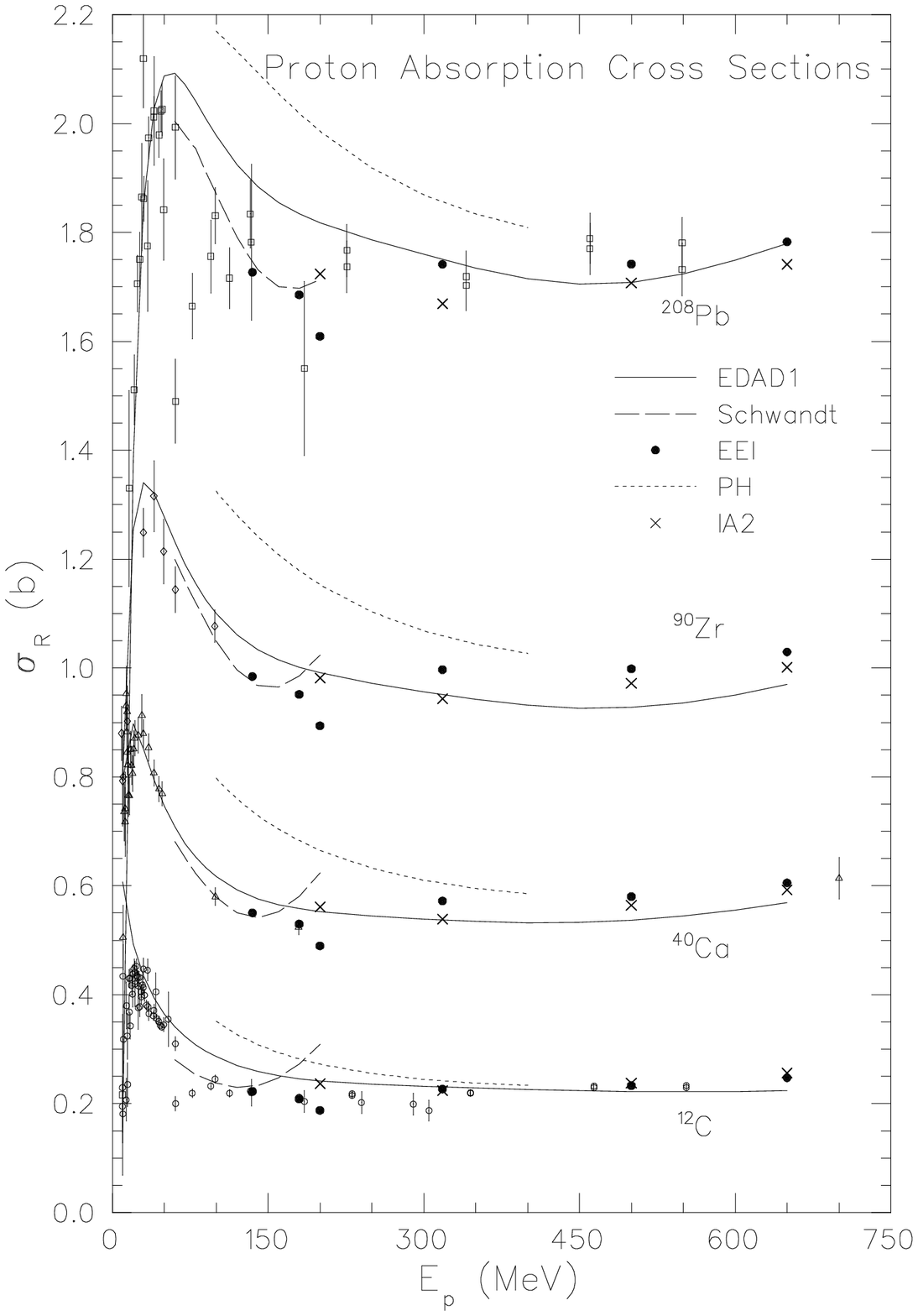,height=6.0in} }
\caption[Proton absorption cross sections.]
{Proton absorption cross sections for several optical models are
compared with data.  
Note that for $^{12}$C and $^{40}$Ca the IA2 and EEI calculations sometimes 
almost coincide and the EEI calculations sometimes obscure data points.}
\label{fig:absorption-xsec}
\end{figure}

Predicted proton absorption cross sections are compared with data in
Fig.\ \ref{fig:absorption-xsec}.
Unfortunately, the available proton data are scarce and of uneven quality 
\cite{Bauhoff86}.
Nevertheless, the EEI model provides accurate predictions for these data,
although the 200 MeV interaction appears to give results which are 
systematically low compared to the trends for other energies.
It is important to remember that the EEI model is dominated by inelastic
data, but gives good fits to elastic data whether or not they are included
in the analysis \cite{Seifert93}.  
Furthermore, neither absorption nor total cross section
data were included in the analysis, but are nevertheless predicted accurately.
Dirac phenomenology also provides good predictions for proton absorption 
cross sections, but its calculations for both $^{12}$C and $^{40}$Ca
appear to be slightly too large.
Below about 150 MeV the Schwandt model also agrees with the data, but its
energy dependence appears to be unreasonable and it begins to diverge from
the data for higher energies.
The earlier version of that model due to Nadasen \cite{Nadasen81} gives
a better description of the energy dependence of the absorption cross section.

\begin{figure}[htbp]
\centerline{
\strut\psfig{file=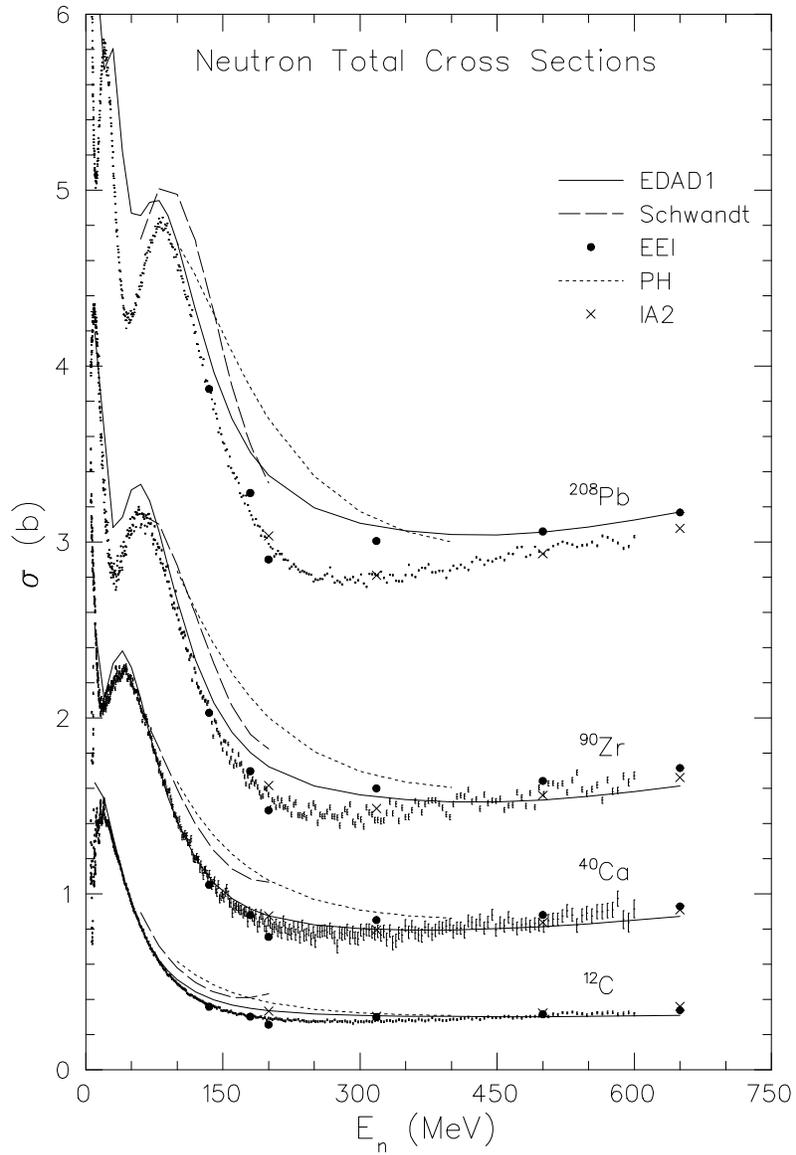,height=6.0in} }
\caption[Neutron total cross sections.]
{Neutron total cross sections for several optical models are
compared with data.
Note that for $^{12}$C and $^{40}$Ca the IA2 and EEI calculations sometimes 
almost coincide.}
\label{fig:ntot}
\end{figure}

In Fig.\ \ref{fig:ntot} predictions for neutron total cross sections 
are compared with the high-quality neutron total cross section data 
recently obtained at LAMPF \cite{Finlay93}.
Neutron total cross sections computed from the Schwandt potential are
substantially larger than the data.
For self-conjugate targets Dirac phenomenology provides good predictions over 
broad ranges of mass and energy, but, lacking a parametrization of the 
symmetry potential, the EDAD1 model fails to reproduce $\sigma_n$ for $N>Z$
\cite{Finlay92}.
As for the proton absorption cross section, EDAD1 predictions for the $^{12}$C  
neutron total cross sections appear to be slightly too large in
the energy range relevant to existing proton knockout data for discrete states.
The EEI model also provides good predictions for self-conjugate targets and
appears to be more accurate than EDAD1 for light nuclei, such as $^{12}$C. 
The result at 200 MeV appears to be slightly too small, as also observed
in the proton absorption cross sections.
Below about below 300 MeV, the EEI calculations for heavier targets with 
significant neutron excesses also appear to be more accurate than EDAD1,
but tend to be too high at higher energies, particularly for $^{208}$Pb.
For these EEI calculations the symmetry potential is obtained by folding 
the isovector density with the density-independent isovector interaction 
from the FL $t$-matrix.
Although the isovector interaction has not been calibrated to 
nucleon-nucleus scattering data with the same care as the isoscalar 
interaction, its contribution is small enough that residual errors in 
theoretical models of that term should not be too serious.
On the other hand, even though the EEI model provides a good fit to proton
elastic scattering by heavy nuclei, there may still be some inaccuracy for
large $A$ because the model was fitted to data for $A \leq 40$.

We find that the IA2 interaction provides accurate predictions for elastic
and inelastic scattering for 500 MeV protons, where the IA2 interaction is
most similar to the empirical effective interaction, 
but that the real--central repulsion of the IA2 model is too strong at lower 
energies.
Nevertheless, Fig.\ \ref{fig:ntot} shows that very accurate predictions are
obtained for neutron total cross sections \cite{Kelly96}.
Similarly, calculations using the relativistic impulse approximation,
with \cite{Kelly94a} or without \cite{Jin93a} density dependence, also
reproduce the the neutron total cross section data for $^{208}$Pb.
Thus, it will be of interest to compare IA2 predictions with measurements of 
nuclear transparency at higher energies soon to be made at CEBAF.

In Figs.\ \ref{fig:absorption-xsec} and \ref{fig:ntot} we also show 
calculations based upon the Paris-Hamburg (PH) interaction.
Although the PH optical potential provides good fits to proton elastic
scattering data \cite{Rikus84,Rikus84b}, 
we find that its predictions for integrated cross sections
are substantially larger than the data.
Similarly, PH calculations for proton inelastic scattering to states with
surface-peaked transition densities also tend to produce cross sections
that are too large \cite{Kelly89a,Seifert93,Kelly91a}.
We have argued \cite{Kelly89b} that for finite nuclei nonlocal corrections to 
the LDA suppress the interaction strength in the surface region, producing
smaller inelastic cross sections for surface-peaked states and smaller
integrated cross sections.
This effect is included in the empirical effective interaction and accounts
in part for the improvement of the EEI with respect to the PH model for these
quantities.
Nevertheless, the PH model provides equal or better fits to proton elastic
scattering.
Thus, there is no guarantee that optical models fitted to elastic scattering 
data alone will provide the best predictions for absorption or total cross 
sections or for nuclear transparency.
Phenomenological models, such as EEI, which include data that are sensitive
to the interior wave function, such as proton inelastic scattering, should
provide more accurate interior potentials and better predictions for
nuclear transparency.
Furthermore, we have also shown that cross sections for exclusive 
$(e,e^\prime p)$ reactions to discrete final states correlate well with 
absorption or total cross sections for nucleon-nucleus scattering 
\cite{Kelly96}.

\section{Results}
\label{sec:results}

\subsection{Comparison with data for $T_p \approx 180$ MeV}
\label{sec:MIT-results}

\begin{figure}[htbp]
\centerline{
\strut\psfig{file=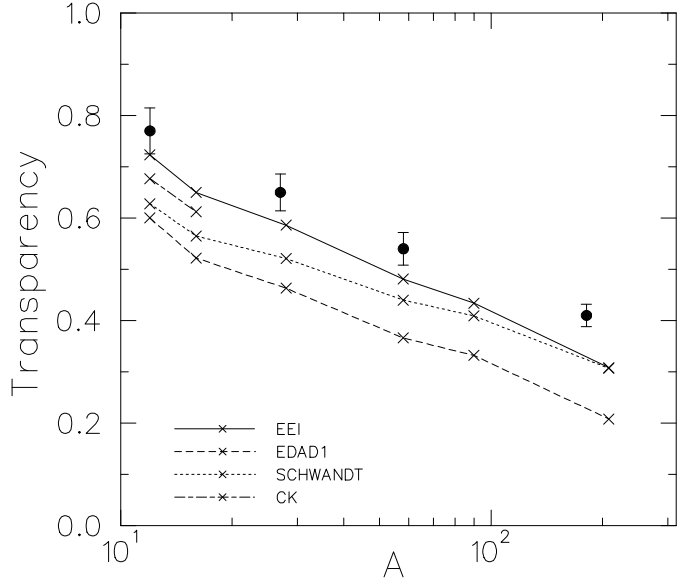,height=3.0in} }
\caption[Nuclear transparency for $T_p = 180$ MeV.]
{Nuclear transparency data for $T_p \approx 180$ MeV are compared with
calculations of ${\cal T}_\bot$ using several optical models.
The data (solid points) were obtained using a 780 MeV electron beam, 
$\omega = 215 \pm 20$ MeV, and $q \approx 605$ MeV/c.
Calculations were performed for selected closed sub-shell nuclei with 
mass numbers indicated by crosses.
The solid line employs the EEI, dashes the EDAD1, dots the Schwandt,
and dash--dots the Comfort and Karp (CK) potential.}
\label{fig:MIT}
\end{figure}

The first measurements of nuclear transparency for $(e,e^\prime p)$ with
$150 \leq T_p \leq 210$ MeV were made by 
Garino {\it et al.} \cite{Garino92,Geesaman89}
using a 780 MeV electron beam, $\omega = 215 \pm 20$ MeV, and 
$q \approx 610$ MeV/c.
Measurements were made for four in-plane opening angles between about 
$0^\circ$ and 23$^\circ$ with $\phi = 180^\circ$, which sample a slice
through the Fermi cone for this $q$.
The ratios
\begin{equation}
  {\cal R}(\theta) = 
 \frac{ \int dE_m \; d^5\sigma(e,e^\prime p) / dE_m d\Omega_e d\Omega_p }
      { d^2\sigma(e,e^\prime) / d\Omega_e }
\end{equation}
integrated over a wide range of missing energy (up to about 100 MeV)
were measured for $^{12}$C, $^{27}$Al, $^{58}$Ni, and $^{181}$Ta.
Several methods were used to relate the measured ratios ${\cal R}(\theta)$
to the nuclear transparency, but the variations were less than 5\%, which
can be viewed as an estimate of the systematic error.
From our point of view, the experimental realization of 
Eq.\ (\ref{eq:perp}) would be
\begin{equation}
\label{eq:transparency-exp}
   {\cal T}_\bot = 
  \frac{  \int d\theta \sin\theta \; {\cal R}^{DW}(\theta) }
       {  \int d\theta \sin\theta \; {\cal R}^{PW}(\theta) } \; ,
\end{equation}
where the range of integration is restricted to the large angle side of
${\bf q}$.
The data for $E_m \leq 80$ MeV are shown in Fig. \ref{fig:MIT}.

Since this experiment was performed with a relatively narrow acceptance
in electron energy,  the calculations were performed with fixed 
$\omega=215$ MeV and $q=605$ MeV/c.
The proton spectrometer, on the other hand, accepted a wide range
of proton energies centered about 180 MeV.
Hence, for each shell the laboratory proton energy for parallel
kinematics was computed as $T_p = \omega - E_m$ where $E_m$ is the 
Hartree--Fock single-particle energy.
The invariant mass for each final state was held constant, so that the
laboratory proton energies for nonparallel kinematics are slightly smaller.
The optical potential for each shell was calculated using the proton
energy for that shell.

Unfortunately, the EEI interaction is only available for discrete energies
and the parameters do not vary as smoothly as one might like because those
interactions were obtained by fitting uncorrelated data sets independently.
No attempt has yet been made to impose a smooth energy dependence upon the
empirical effective interaction.
Nevertheless, the interactions do vary smoothly enough that calculations
made using the 180 MeV or the 200 MeV interaction are quite similar and the
small differences between them do not affect the conclusions.
Furthermore, a large part of the energy dependence of the effective 
interaction arises from the exchange contributions, which were evaluated
using the kinematics of each orbital.  
Therefore, we chose to use the fit to data for 180 MeV protons, representing 
the center of the proton energy distribution.  
Similar calculations using the 200 MeV EEI interaction, fixed $\omega$
rather than fixed $T_p$, and a other few technical differences, 
were presented in Ref.\ \cite{Kelly96}.
Although slightly better agreement with the data was obtained using that
slightly less absorptive interaction, the small difference between the two
calculations demonstrates that the results are relatively insensitive to
uncertainties in the EEI parametrization and ambiguities in the prescription 
for acceptance averaging.

We have studied the sensitivity of nuclear transparency to the optical model 
by comparing the EEI model with Dirac phenomenology, version EDAD1. 
Although the data extend in both energy and mass beyond their ranges of
applicability, we also show calculations based upon the Schwandt and
the Comfort and Karp (CK) optical potentials for comparison with other 
authors who have used those models.
Ejectile wave functions for the Schwandt and CK potentials include  
Perey factors, Eq. (\ref{eq:PB}), with $\beta = 0.85$ fm, 
whereas wave functions based upon Schr\"{o}dinger--equivalent potentials 
from Dirac phenomenology include the Darwin factor, Eq. (\ref{eq:Darwin}).
Nonlocality corrections are not needed for the EEI model because both
elastic and inelastic scattering are fitted self-consistently in that 
model.
These calculations are compared with the  data in 
Fig. \ref{fig:MIT}.
We find that the EEI model provides a good description of nuclear 
transparency, whereas considerably smaller transparencies are obtained
with either the Schwandt or the EDAD1 potentials.
These results are consistent with the observation in 
Sec.\ \ref{sec:comparison-of-potentials}
that the EEI model provides more accurate predictions for proton
absorption and neutron total cross sections also.
It is clear that the EDAD1 model is too absorptive at these energies, so
that spectroscopic factors using it will be artificially large to compensate
for excessive attenuation.

Similar calculations have been performed by 
Ireland {\it et al.} \cite{Ireland94} using the Schwandt potential and
they obtained larger transparencies which agree better with the data than 
do our ostensibly similar calculations with the same potential.    
However, that analysis suffers from several defects.
First, the kinematics were artificially altered so that knockout from every
orbital was assigned the same ejectile energy, 180 MeV, despite the
wide range of energies covered by the experiment.
Thus, they used the Schwandt potential at 180 MeV even though, for the
same electron kinematics, protons ejected from less bound orbitals actually
emerged with energies beyond the range of that optical model.
Second, their calculations used a scattered electron energy that is 30 MeV 
higher than the center of the experimental acceptance.
Third, the DWIA cross section was computed using a current operator based
upon nonrelativistic reduction to second order in $q/m_N$,
but the distorted momentum distribution was obtained by dividing the cross
section by $\sigma_{cc1}$, the cross section for the {\it cc1} current operator.
The inconsistency between the numerator and the denominator in their
application of Eq.\ (\ref{eq:transparency}) leads to a spurious enhancement of
the transparency above unity for a plane--wave calculation.
However, none of these defects appears to be sufficient to explain the
discrepancy between the two calculations.
Nevertheless, we consider the apparent agreement between the data
and the Schwandt calculations of Ref.\ \cite{Ireland94} to be a fortuitous 
result of an incorrect calculation.

Phenomenological optical models, nonrelativistic or relativistic, which are
fitted only to elastic scattering data may fit that data well but still
fail to predict nuclear transparency correctly because such analyses are
not sensitive to the interior wave function.
The EEI model is much more sensitive to the interior wave function because
in fitting inelastic scattering data it requires consistency between 
distorted waves and inelastic transition amplitudes.
Therefore, the EEI model provides a more accurate prediction of nuclear
transparency and should also provide more accurate spectroscopic factors.
Clearly it will be of interest to obtain transparency data for a wider
range of energies and such studies are planned for CEBAF.
It would also be of interest to obtain comparable data at lower energies
where NIKHEF has performed its extensive survey of spectroscopic factors
for complex nuclei \cite{Kelly96}, 
but it appears that the lower--energy regime will soon be abandoned.

\subsection{Comparison with data for $T_p \approx 650$ MeV/c}
\label{sec:SLAC-results}

\begin{figure}[htbp]
\centerline{
\strut\psfig{file=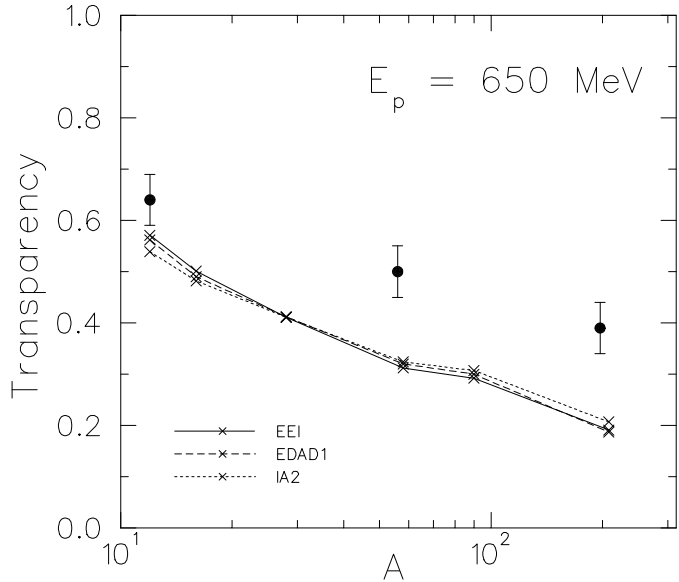,height=3.0in} }
\caption[Nuclear transparency for $T_p = 650$ MeV.]
{Nuclear transparency data for $Q^2 = 1.04$ (GeV/c)$^2$ are compared with
calculations of ${\cal T}_\bot$ using several optical models at $T_p = 650$ MeV.
Calculations were performed for selected closed sub-shell nuclei with 
mass numbers indicated by crosses.
The solid line employs the EEI, dashes the EDAD1, dots the IA2 optical
models.}
\label{fig:SLAC}
\end{figure}

Nuclear transparency data for $1.0 \leq Q^2 \leq 6.8$ (GeV/c)$^2$ have
recently been obtained at SLAC by the NE18 collaboration
\cite{Makins94,O'Neill95}.
In Fig.\ \ref{fig:SLAC} we compare calculations with the data for
$Q^2 = 1.04$ (GeV/c)$^2$ and $\omega = 0.625$ GeV.
The angular integrations were symmetric with respect to ${\bf q}$.
Since the dependence of nuclear transparency on proton energy is quite
slow for these energies (see following section), we performed the calculations
using a fixed proton energy of 650 MeV for which both the EEI and IA2 
interactions are available.
The electron kinematics were then computed using $\omega = E_m + T_p$ for
each shell.
At this energy we find that the EEI, IA2, and EDAD1 potentials all give
practically the same results --- the variation due to choice optical potential
is much smaller at 650 than at 180 MeV.
However, although the calculations are fairly close to the data for $^{12}$C,
the data vary much less with $A$ than do the calculations.
O'Neill {\it et al}.\ report that the data can be fitted with a power law
of the form $A^{-\alpha}$ with $\alpha = 0.18$ at $Q^2 = 1.0$ (GeV/c)$^2$
and with slightly larger values of $\alpha$ at higher $Q^2$.
However, the calculations are much closer to the characteristic 
$\alpha = 1/3$ behavior that would be expected for an eikonal model 
with constant attenuation length.

The experimental values of ${\cal T}_\bot$ reported by the NE18
collaboration \cite{Makins94,O'Neill95} include corrections for the 
portion of the model spectral functions which would fall outside their
acceptances.
These correction factors are 1.11, 1.22, and 1.28 for carbon, iron, and gold, 
respectively.
The calculations of Nikolaev {\it et al}.\ \cite{Nikolaev94b} appear to be
in good agreement with the $A$-dependence of the data for $Q^2 \approx 1.0$
(GeV/c)$^2$, but their figure apparently employs a preliminary analysis of
the data or neglects these correction factors resulting in smaller
experimental values.
However, elimination of the correction factors would not be enough
to bring the data into agreement with our calculations.
As discussed in Sec. \ref{sec:discussion}, the most important difference
between our calculations and those of Nikolaev {\it et al}.\ can be
traced to differences between the definitions of the semi-inclusive
cross section from which transparencies are calculated.
Nevertheless, it would be of interest to compare proton absorption and neutron
total cross sections computed with their Glauber model with nucleon-nucleus
data for this energy regime.

\subsection{Energy Dependence of Nuclear Transparency}
\label{sec:energy-dependence}

\begin{figure}[htbp]
\centerline{
\strut\psfig{file=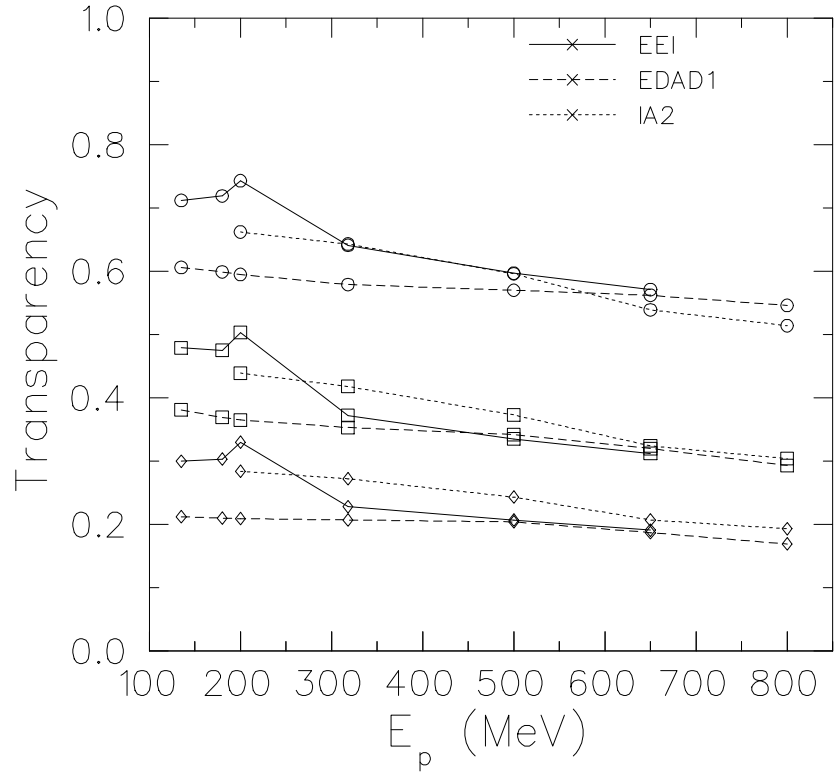,height=4.0in} }
\caption[Energy dependence of nuclear transparency.]
{The energy dependence of nuclear transparency is shown for $^{12}$C (circles),
$^{58}$Ni (squares), and $^{208}$Pb (diamonds) using quasiperpendicular
kinematics for 2.1 GeV electrons.
Calculations are shown for the EEI, EDAD1, and IA2 models as solid, dashed, 
and dotted lines, respectively.}
\label{fig:energy-dependence}
\end{figure}

The energy dependence of nuclear transparency is examined in 
Fig.\ \ref{fig:energy-dependence},
which compares calculations using the EEI, IA2, and EDAD1 potentials for
representative light ($^{12}$C), medium ($^{58}$Ni), and heavy ($^{208}$Pb)
nuclei.  
The calculations were performed for proton energies  marked by symbols, 
where circles are for $^{12}$C, squares are for $^{58}$Ni, and diamonds are
for $^{208}$Pb.
The calculations were made for constant $(\omega,q)$ kinematics using a beam 
energy of 2.1 GeV, although the results are practically independent of the 
electron energy.
The energy transfer for each shell was taken to be $\omega = E_m + T_p$,
where $T_p$ is the laboratory kinetic energy for zero recoil and $E_m$ is
the Hartree--Fock energy for the shell.
The angular integrations were symmetric with respect to ${\bf q}$.

The proton energies were selected according to the availability of EEI and/or 
IA2 interactions.
The small kinks in the EEI calculations between about 180 and 200 MeV
probably indicate the degree of model dependence that arises when fitting
independent data sets without imposing a smooth energy dependence.
These kinks are clearly correlated with the corresponding neutron total
cross section calculations shown in Fig.\ \ref{fig:ntot}, which suggest that
the 200 MeV empirical effective interaction is not quite sufficiently
absorptive.
On the other hand, the EEI calculations of neutron total cross sections for
heavy nuclei tend to be a little too high at 180 MeV.
Hence, the best estimate of the transparency between 180 and 200 MeV
probably lies between the calculations shown in 
Fig.\ \ref{fig:energy-dependence}, with an uncertainty comparable to the
difference between them.

For energies below about 200 MeV, the EEI model is more transparent than
any of the other optical models we have examined and provides the most
accurate predictions for the MIT data.
Nevertheless, those data still remain about 5--10\% above the EEI calculations.
At higher energies the variation among optical potentials tends to decrease,
although for medium and heavy nuclei the IA2 calculations are more transparent
than either the EEI or EDAD1 models.
This behavior is also clearly correlated with the neutron total cross section
calculations shown in Fig.\ \ref{fig:ntot}, where for medium and heavy
nuclei the IA2 interaction provides accurate predictions while the EDAD1
and EEI calculations are similar and are both slightly too large.

A series of $^{12}$C$(e,e^\prime p)$ experiments performed at MIT-Bates
in parallel kinematics near the quasifree ridge between  
$0.14 \lesssim Q^2 \lesssim 0.83$ (GeV/c)$^2$ shows that there is a significant,
nearly constant, continuum yield extending to very large missing energies
even when the missing momentum is relatively small.
The continuum yield at large $E_m$ is predominantly transverse and cannot be 
reproduced by calculations of multiple scattering in the final state.
A review of these data may be found in Ref.\ \cite{Kelly96}.
One possible interpretation of these results is that multinucleon absorption 
of the virtual photon enhances the yield at large missing energies even when 
the missing momentum is relatively small.
Since the present model does not include multinucleon absorption mechanisms,
perhaps it is not surprising that DWA calculations underestimate the
semi-inclusive cross section for $^{12}$C$(e,e^\prime p)$.
For $\omega=215$ MeV the ratio between experiment and calculation is
approximately constant, but for $\omega \approx 650$ MeV, the discrepancy
increases with $A$.
Although no calculations of multinucleon absorption are available for
these kinematics, it seems reasonable to assume that such effects can become
more important as either $A$ or $\omega$ increases such that more energy
is available to be shared among more nucleons at greater average density.
Furthermore, Lourie {\it et al}.\ \cite{Lourie93} have suggested that
kinematic focussing of the multinucleon phase space would exacerbate the
artificial enhancement of semi-inclusive transparency measurements
as $Q^2$ increases.

\section{Discussion}
\label{sec:discussion}

Most previous calculations of nuclear transparency have been based upon
eikonal models.
For example, Pandharipande and Pieper \cite{Pandharipande92}
studied nuclear transparency using a correlated Glauber model in which
\begin{mathletters}
\label{eq:Pieper}
\begin{eqnarray}
  {\cal T} &=& \frac{1}{Z} \int d^3r^\prime \, 
  \rho_p({\bf r}^\prime) P_T({\bf r}^\prime) \\
 P_T({\bf r}^\prime) &=& \exp\left\{ -\int_{z^\prime}^\infty dz^{\prime \prime}
 \left[ g_{pn}({\bf r}^\prime,{\bf r}^{\prime\prime}) 
        \tilde{\sigma}_{pn}(q,\rho({\bf r}^{\prime \prime}))
        \rho_n({\bf r}^{\prime \prime}) 
      + g_{pp}({\bf r}^\prime,{\bf r}^{\prime\prime})
        \tilde{\sigma}_{pp}(q,\rho({\bf r}^{\prime \prime}))
         \rho_p({\bf r}^{\prime \prime}) \right] \right\}
\end{eqnarray}
\end{mathletters}
where $P_T({\bf r}^\prime)$ represents the probability that a proton
struck at position ${\bf r}^\prime$ will emerge without rescattering,
$\tilde{\sigma}_{pN}(q,\rho({\bf r}^{\prime \prime}))$ represents the
effective $pN$ cross section evaluated in local density approximation,
and 
\begin{displaymath}
 g_{pN}({\bf r}^\prime,{\bf r}^{\prime\prime}) \approx
   g_{pN}(\rho_0, |{\bf r}^\prime - {\bf r}^{\prime\prime}|)
\end{displaymath}
represents the pair distribution function evaluated at central density.
Although this calculation agrees well with the MIT data, it should be 
noted that Eq. (\ref{eq:Pieper}) applies to ${\cal T}_\|$ whereas the
experiment measured ${\cal T}_\bot$.
It is also important to recognize that despite the differences between the
formulations of these models, the EEI model includes essentially the same 
physics.
Recall that the EEI parametrization was originally based upon 
Brueckner--Hartree--Fock calculations of the effective interaction in 
nuclear matter which include both short--range correlations and Pauli
blocking, but that the parameters were adjusted to improve the fit
to experimental data.
The correlated Glauber model includes Pauli blocking in its effective cross 
section, whereas the EEI model includes it as a density-dependent damping of 
its imaginary central interaction.
Similarly, the anticorrelation between identical nucleons is represented
by the pair distribution function in the correlated Glauber model or by
the density--dependent short-ranged repulsion in the real central EEI
interaction.
Pandharipande and Pieper find that Pauli blocking, effective mass, and
correlation effects all play important roles in their transparency
calculations.

However the validity of the Glauber model is questionable at 
low energies where rectilinear propagation is a poor approximation, 
particularly for energies as low as 180 MeV used in the MIT experiment
and the corresponding calculations of Ref.\ \cite{Pandharipande92}.
The Glauber approximation also uses a zero-range approximation to the 
nucleon-nucleon interaction, which does not apply at intermediate energies
even if Pauli blocking and correlation corrections are made.
The local density approximation provides a more realistic description of the
radial dependence of the optical potential, 
but includes the Glauber approximation as a limiting case.
For proton elastic scattering the Glauber model is generally considered 
adequate for $T_p \gtrsim 800$ MeV \cite{Bassichis71,Frankfurt94}.
Although integrated quantities, such as nuclear transparency, are probably
less sensitive to the details of final state interactions, the lowest 
energy at which the Glauber approximation to nuclear transparency can be
employed has not been established.
On the other hand, the Glauber approximation is more efficient than the 
distorted wave approximation and should become sufficiently accurate 
for $T_p \gtrsim 800$ MeV. 
Furthermore, we find that the variation among optical potentials also
decreases at higher energies where the Glauber approximation becomes
applicable.
Therefore, Glauber calculations are more appropriate for 
$Q^2 \gtrsim 2.0$ (GeV/c)$^2$.

Neither the optical nor the Glauber model consider in detail the distribution
of absorbed flux among final states of the residual system.
To the extent that nucleon knockout dominates the absorption cross section
at intermediate energies, most of the flux leaving the elastic channel
leads to multinucleon final states in the continuum which should be
excluded from the semi-inclusive cross section used to measure nuclear
transparency.
Of course, it is much easier to exclude the multinucleon continuum 
theoretically than experimentally.
Inelastic processes within final-state interactions that result in single-hole
states of the residual nucleus would require a coupled-channels model for
detailed analysis and should be included in the semi-inclusive cross section,
but represent a rather small fraction of the absorption cross section.
Therefore, we expect that most inelastic processes reduce transparency as
stipulated by the optical model and the distorted wave approximation.

Nuclear transparency for $(e,e^\prime p)$ at large $Q^2$ is also usually 
interpreted using a Glauber model.
As developed by Nikolaev {\it et al.}\ \cite{Nikolaev95a,Nikolaev95b}, 
the Glauber transparency functions for parallel and quasiperpendicular 
kinematics take the form
\begin{mathletters}
\label{eq:Nikolaev}
\begin{eqnarray}
{\cal T}^G_\| &=& \frac{1}{A \sigma_{tot}(pN)} \int d^2b \; \left\{
1 - \exp \left[- \sigma_{tot}(pN) t(b) \right] \right\} 
\label{eq:Nikolaev-para} \\
{\cal T}^G_\bot &=& \frac{1}{A \sigma_{in}(pN)} \int d^2b \; \left\{
1 - \exp \left[- \sigma_{in}(pN) t(b) \right] \right\}
\label{eq:Nikolaev-perp}
\end{eqnarray}
\end{mathletters}
where 
\begin{equation}
 t(b) = \int_{-\infty}^{\infty} dz \; n_A(b,z)
\end{equation}
is the optical thickness at impact parameter $b$ through nuclear density
$n_A(b,z)$.
These results for nuclear transparency in the Glauber model are distinguished
from our definitions by the superscript $G$.
The transparency for parallel kinematics, ${\cal T}^G_\|$, is governed by the 
total proton-nucleon cross section, $\sigma_{tot}(pN)$, which represents the
loss of flux in the forward direction (parallel to the momentum transfer).
The transparency for quasiperpendicular kinematics is larger because 
integration over the full angular range recaptures the part of the flux
that corresponds to elastic proton-nucleon final-state interactions and hence 
${\cal T}^G_\bot$ is governed by the inelastic proton-nucleon cross section, 
$\sigma_{in}(pN)$.
 
Nikolaev {\it et al}.\ \cite{Nikolaev94b} demonstrated that the difference
between ${\cal T}^G_\|$ and ${\cal T}^G_\bot$ is quite large at low $Q^2$
where the inelasticity of the nucleon-nucleon interaction is relatively small.
Furthermore, averaging over the acceptance for experiment NE18, they obtained
an accurate description of the data that is intermediate between 
${\cal T}^G_\bot$ and ${\cal T}^G_\|$.
However, for $Q^2 < 1$ (GeV/c)$^2$ the difference between 
${\cal T}_\|$ and ${\cal T}_\bot$ 
is much smaller in the distorted wave approximation than in the Glauber model
(see the end of Sec.\ \ref{sec:definition}).
Since the absorptive content of the optical potential
for energies below the pion production threshold is determined by 
quasifree elastic nucleon-nucleon scattering, one expects  
${\cal T}^G_\| \approx {\cal T}_\| \approx {\cal T}_\bot$ for energies
large enough to neglect Pauli blocking.
On the other hand, 
the most important reason why the Glauber model calculations of 
Nikolaev {\it et al}.\ \cite{Nikolaev94b}
produce larger transparencies than our optical models for 
$Q^2 \gtrsim 1$ (GeV/c)$^2$ can be traced to differences between the
definitions of the semi-inclusive cross sections used to derive 
${\cal T}^G_\bot$ and ${\cal T}_\bot$ that arise from differences between
a high-energy versus a low-energy viewpoint.

The derivation of Eq. (\ref{eq:Nikolaev-perp}) requires a closure sum over
all final states of the residual nuclear system that contain $A-1$ nucleons
({\it e.g.} Ref. \cite{Nikolaev95b}).
Thus, the semi-inclusive cross section used for  ${\cal T}^G_\bot$ 
includes processes in which one or more secondary nucleons is
ejected from the residual nucleus by final-state interactions with the
primary ejectile even if the resultant missing energy is outside the 
experimental acceptance.
From a high-energy viewpoint, such processes are described as incoherent
elastic rescatterings of the ejectile and are driven by the elastic
proton-nucleon cross section, $\sigma_{el}(pN)$. 
In fact, the only final states excluded from the semi-inclusive cross
section for ${\cal T}^G_\bot$ are those which contain additional particles,
typically one or more pions for modest $\omega$.
Therefore, in the absence of Fermi motion, ${\cal T}^G_\bot$ for ejectile
energies below the pion production threshold would be unity because the
proton-nucleon interaction is then elastic.
According to Table 1 of Ref.\ \cite{Nikolaev95b}, 
${\cal T}^G_\bot \approx 0.97$ for $^{12}$C or 
${\cal T}^G_\bot \approx 0.84$ for $^{208}$Pb
remains large even for $Q^2 \sim 1$ (GeV/c)$^2$.
For larger ejectile energies ${\cal T}^G_\bot$ decreases as the
proton-nucleon interaction becomes increasing inelastic and approaches
${\cal T}^G_\|$ for large $Q^2$.

By contrast, the semi-inclusive cross section used to obtain ${\cal T}_\bot$ 
within the optical model excludes multinucleon emission arising from
final-state interactions.
From a low-energy viewpoint, nuclear inelastic scattering which changes the
internal state of the residual nucleus is described as a loss of flux from
the elastic channel.
Thus, inelastic scattering involving a small energy transfer between the
ejectile and the residual contributes to the absorptive (imaginary) potential
even if the final state remains a low-lying excitation of the $(A-1)$-system.
Since some of these low-lying final states fall within the missing-energy
acceptance and are thereby included in the experimental semi-inclusive 
cross section, optical model calculations using potentials fitted to 
elastic nucleon-nucleus scattering probably slightly overestimate the 
loss of flux.
More accurate models which account for this effect would require
coupled-channels calculations based upon optical potentials constructed for
the model space included within the semi-inclusive cross section, but,
fortunately, since the imaginary part of the optical potential for 
intermediate energy nucleons with $100 \lesssim T_p \lesssim 800$ MeV is 
dominated by quasifree knockout processes, this error is small and decreases 
as the ejectile energy increases.
Nevertheless, it is clear that the more restrictive summation over final
states in the optical model definition of nuclear transparency must result
in ${\cal T}_\bot < {\cal T}^G_\bot$.
Furthermore, the optical model definition of nuclear transparency is more
relevant to exclusive $(e,e^\prime p)$ measurements of the missing-momentum
distributions for discrete final states and to semi-inclusive experiments
of the type performed by MIT and illustrated in Fig.\ \ref{fig:MIT},
for which the Glauber model would give a result, ${\cal T}^G_\bot \approx 1$,
that is much larger than the data because the ejectile energy is below
the nucleon-nucleon inelasticity threshold.
Although several other assumptions employed in its derivation also
fail and preclude its application below about 1 (GeV/c)$^2$,
the most important deficiency of this version the Glauber model for low $Q^2$ 
is that its summation over final states is much more inclusive than the
experimental definition of the semi-inclusive cross section.

The Glauber model for ${\cal T}^G_\bot$ emphasizes the inelasticity of the
nucleon-nucleon interaction, whereas the optical model for ${\cal T}_\bot$ 
emphasizes the inelasticity of the nucleon-nucleus interaction.
Those final-state interactions which eject one or more low-energy
nucleons but leave the missing energy within the integration range of
the experiment reduce ${\cal T}_\bot$ but do not reduce ${\cal T}^{exp}_\bot$.
On the other hand, elastic proton-nucleon interactions which increase
the missing energy too much reduce ${\cal T}^{exp}_\bot$ but do not reduce
${\cal T}^{G}_\bot$.
Therefore, the experimental definition of the semi-inclusive cross section for
$(e,e^\prime p)$ reactions lies between those used for ${\cal T}^G_\bot$ 
and ${\cal T}_\bot$.
For relatively low $Q^2$ the optical model is most appropriate, but will
generally underestimate the transparency whereas the Glauber model for
${\cal T}_\bot$ will substantially overestimate the transparency.
For large $Q^2$ where the reactive content of the optical potential is
dominated by nucleon-nucleon inelasticity we would expect ${\cal T}_\bot$
to approach ${\cal T}^G_\bot$ from below, but the Glauber model is clearly
much more efficient computationally than the optical model under these
conditions.
In between we would expect these two models to bracket the data.
However, neither approach includes multinucleon absorption of the virtual
photon upon a correlated cluster, which also might increase the experimental
cross section with respect to direct knockout models.

Experimentally it would be interest to observe multinucleon knockout by
electron scattering and to study the kinematic dependencies of various
multinucleon channels in detail.
Although it is not possible to separate the various processes which lead
to the same final state in a model independent fashion, the kinematic 
differences among them can be usefully analyzed in the context of a model.
For example, single-nucleon knockout from a deeply bound orbital
leaves the residual nucleus in a highly excited state which may decay by 
particle emission and hence produce multinucleon emission without need
of final-state interactions.
Multinucleon processes of this type are probably close to isotropic in
the rest frame of the residual nucleus and from an optical-model viewpoint
should be included in the semi-inclusive cross section because
final-state interactions are not required.
On the other hand, one might expect secondary nucleons knocked out by
interactions with the ejectile to appear preferentially in the forward
hemisphere; such events should be included in the cross section for
${\cal T}^G_\bot$ but excluded for ${\cal T}_\bot$.
Analyses of this type require measurement of the angular correlation
between the high-energy primary proton and one or more low-energy 
secondary nucleons.
Similarly, although it is not possible to unambiguously identify events
arising from multinucleon absorption of the virtual photon upon a 
correlated cluster, an enhancement of the probability for observing 
two nucleons corresponding to missing momenta $\pm p_m$ 
with $p_m \gtrsim k_F$ can be interpreted within a model as a signature
of short-range correlations.
Although one would expect the longitudinal/transverse character of 
multinucleon emission arising from the decay of a deep-hole state or from 
final-state interactions to remain consistent with the direct knockout
model, multinucleon absorption of the virtual photon could substantially
alter the structure of the response functions.
In fact, there is some evidence at low $Q^2$ that nucleon knockout at
large missing energy is enhanced by a process, perhaps multinucleon absorption,
that is largely transverse.
However, clarification of these issues will require considerably more
work, both experimental and theoretical.

It may be possible to refine optical model calculations of nuclear
transparency by applying the statistical multistep direct reaction theory
of Feshbach, Kerman, and Koonin \cite{Feshbach80} to evaluate the energy
and angular distributions of protons which suffer final state interactions
and thereby to estimate the fraction of the flux described by the optical
model as absorption that actually remains within the experimental acceptances.
However, it would then also be necessary to test those calculations
against inclusive data for $(p,p^\prime)$.
Such data is available at low energies, 
{\it e.g.} Refs. \cite{Cowley91,Fortsch91},
but is not available for $T_p > 200$ MeV.
If data were available, it should also be possible to estimate the
necessary corrections using a convolution procedure.
However, these possibilities lie well beyond the scope of the present work.

\section{Conclusions}
\label{sec:conclusions}

We have used the distorted wave approximation to evaluate nuclear transparency
to intermediate-energy protons in semi-inclusive $(e,e^\prime p)$ reactions.
We compared calculations using density-dependent effective interactions
from the EEI model, 
which is fitted to proton elastic and inelastic scattering data,
and the IA2 model, which is derived from a relativistic boson exchange model,
with global optical potentials from Dirac phenomenology.
For low energies we also considered several traditional nonrelativistic
optical models.
We demonstrated that nuclear transparency in $(e,e^\prime p)$ reactions
is well correlated with the proton absorption and neutron total cross sections 
calculated using these models.
The IA2 model was found to give the most accurate predictions for neutron
total cross sections at $T_p \gtrsim 200$ MeV.
The EEI model also provides accurate predictions and extends to lower energies, 
but slightly overestimates the neutron total cross sections for
heavy nuclei at $T_p \gtrsim 300$ MeV.

For ejectile energies near 200 MeV we find that there is considerable
sensitivity to the choice of optical model.
Global optical potentials from Dirac phenomenology yield nuclear 
transparencies that are much smaller than the data, and hence are likely to 
overestimate spectroscopic factors for discrete states.
Larger transparencies are obtained from the IA2 model, but those calculations
remain significantly lower than the data.
Calculations using the EEI model predict larger transparencies than any other 
model considered, but still remain 5--10\% below the data.
The EEI model predicts greater transparency than other models for 
$T_p \lesssim 200$ MeV, but low-energy semi-inclusive data are presently 
lacking.
At larger ejectile energies the sensitivity to the choice of optical model
is reduced, with all models considered producing similar results for
$T_p \gtrsim 500$ MeV.
Nevertheless, the calculated nuclear transparencies remain substantially
below the NE18 data for $Q^2 \approx 1$ (GeV/c)$^2$, with the discrepancy
increasing with $A$.
Multinucleon contributions to the continuum may enhance the semi-inclusive 
cross section for knockout and thereby artficially enhance the measured 
nuclear transparency.
These contributions are likely to increase with both mass and energy,
but quantitative estimates are not available.

The present model is well suited to the investigation of intermediate-energy
proton knockout to discrete states of the residual nucleus.
The dependence of attenuation factors for valence orbitals upon ejectile
energy can then be used to investigate nuclear transparency, 
where differing radial localizations can help discriminate between
interior and surface properties of the optical potential.
The asymmetry between parallel and antiparallel kinematics for individual
orbitals can be investigated also.
Recoil polarization may also provide additional insight into the final-state 
interactions, and in particular may help to discriminate between
single-nucleon and multinucleon contributions to the continuum.

We have also examined in some detail the difference between Glauber and
optical model calculations of nuclear transparency for quasiperpendicular
kinematics.
By using a much more inclusive summation over final states, 
the Glauber model emphasizes the inelasticity of the nucleon-nucleon 
interaction, whereas with a more restrictive definition of the semi-inclusive
cross section the optical model emphasizes the role of nucleon-nucleus 
inelasticity. 
Therefore, the Glauber model produces larger transparency factors than the
optical model, with the difference between the two approaches becoming
quite large for $Q^2 < 1$ (GeV/c)$^2$.
However, the experimental definition of the semi-inclusive cross section
usually lies between these extremes.
Although the optical model is expected to underestimate the experimental
semi-inclusive cross section, it is much more appropriate for low $Q^2$
than the Glauber model.
For large $Q^2$ where nucleon-nucleon inelasticity accounts for a much larger
fraction of the nucleon-nucleus inelasticity, the two models should produce
similar results, but the Glauber approximation is computationally much more
efficient.

\acknowledgements

The support of the U.S. National Science Foundation under grant PHY-9220690 
is gratefully acknowledged.


\newpage

\end{document}